\documentclass[proceedings, preprint]{rmaa}

\usepackage{paralist}

\usepackage{psfrag,color}

\newcommand{\msun}{$M_{\sun}$\,}

\newcommand{\mbylk}{${M/L_{K}}$}

\SetYear{2009}
\SetConfTitle{A Long Walk Through Astronomy: Luis Carrasco's 60th Birthday}

\title{M82 as a Galaxy: Morphology and Stellar Content of the Disk and Halo}

\author{
  Y. Divakara Mayya\altaffilmark{1}
and Luis Carrasco\altaffilmark{1}
}

\altaffiltext{1}{Instituto Nacional de Astrof\'\i{}sica \'Optica y 
Electr\'onica, Luis Enrique Erro 1, Tonantzintla, C.P. 72840, Puebla, 
M\'exico}
\shortauthor{Mayya}
\shorttitle{M82 as a galaxy}

\listofauthors{Y.D. Mayya \& L. Carrasco}
\indexauthor{Mayya, Y.D.}
\indexauthor{Carrasco, L.}

\abstract{
For decades, the nuclear starburst has taken all the limelight in M82 with
very little discussion on M82 as a galaxy. The situation is changing over
the last decade, with the publication of some important results on the
morphology and stellar content of its disk and halo.
In this review, we 
discuss these recent findings in the framework of M82 as a galaxy.
It is known for almost half a century that M82 as a galaxy doesn't follow
the trends expected for normal galaxies that had prompted the morphologists 
to introduce a separate morphological type under the name Irr II or amorphous. 
It is now being understood that
the main reasons behind its apparently distinct morphological appearance
are its peculiar star formation history, radial distribution of gas density
and the form of the rotation curve. The disk formed 
almost all of its stars through a burst mode around 500 Myr ago, with
the disk star formation completely quenched around 100~Myr ago.
The fossil record of the disk-wide burst lies
in the form of hundreds of compact star clusters, similar in mass to
that of the globular clusters in the Milky Way, but an order of magnitude
younger.
The present star formation is restricted entirely to the central 500~pc 
zone, that contains more than 200 young compact star clusters.
The disk contains a non-star-forming spiral arm, 
hidden from the optical view by a combination of extinction and high
inclination to the line of sight. 
The halo of M82 is also unusual in its stellar content, with
evidence for star formation, albeit at low levels, occurring continuously
for over a gigayear. We carefully 
examine each of the observed abnormality to investigate the overall 
effect of interaction on the evolution of M82.
}

\addkeyword{galaxies: individual (M82) --- galaxies: star clusters}
\addkeyword{galaxies: starburst --- galaxies: stellar content}
\addkeyword{galaxies: structure}

\begin{document}
\maketitle

\section{Introduction}
\label{sec:intro}

Located at only a distance of 3.63 Mpc (image scale 17.6~pc\,arcsec$^{-1}$; 
Freedman et al. 1994), M82 is one of the most observed objects. 
The nuclear starburst and the associated galactic superwinds along
with the filamentary distribution of dust, not only give it a
visually appealing morphology, but also make it an interesting object
of study in the entire range of electromagnetic spectrum.
It is one of the brightest infrared objects in the sky. 
Historically, M\,82 was given a morphological class of Irr II by
\citet{Holm50}, I0 by \citet{deVa59} and ``Amorphous'' by \citet{Sand79}.
It was referred to as an exploding galaxy by \citet{San64} 
due to the discovery of high velocity H$\alpha$ filaments that were thought
to be ejected from the nucleus. 

The exotic idea on the nature of M82 were put to rest by the classical 
works of \citet{Oco78} and \citet{Riek80}, the latter successfully 
explaining all 
the important observed features under the framework of a starburst model.
Since then, the nuclear region of M82 is considered an archetype for 
the starburst phenomenon. The discovery of a bridge of
intergalactic gas connecting M82 to its neighbor M\,81 and subsequent
N-body simulations suggest that the starburst in M82 is triggered by
an encounter between M82 and NGC3077, both members of the M\,81 group
\citep{Gott77, Yun93}.

Though there exist innumerable studies on M82, a vast majority of them 
discuss the nuclear starburst or a phenomenon related to the starburst 
activity. However, interest in M82 as a galaxy is growing in recent years.
There are suggestions that in M82, we may be witnessing the formation
of a new galaxy in our neighborhood.
The principal aim of this review is to summarize the known properties of M82
and discuss these properties in the framework of M82 as a galaxy.

In \S2, we briefly discuss the properties of the central kiloparsec,
which includes the starburst nucleus and the bar. 
The properties of the inner stellar disk (1--3~kpc) are discussed in \S\S3
and 4. The outer disk and the halo are discussed in \S5.
In \S6, we discuss the significance of the presence of
a population of compact star clusters in the disk.
M82 as a galaxy and its current evolutionary status are discussed in \S7. 
Concluding remarks are given in \S8.

\begin{figure*}[!t]
  \includegraphics[width=17cm]{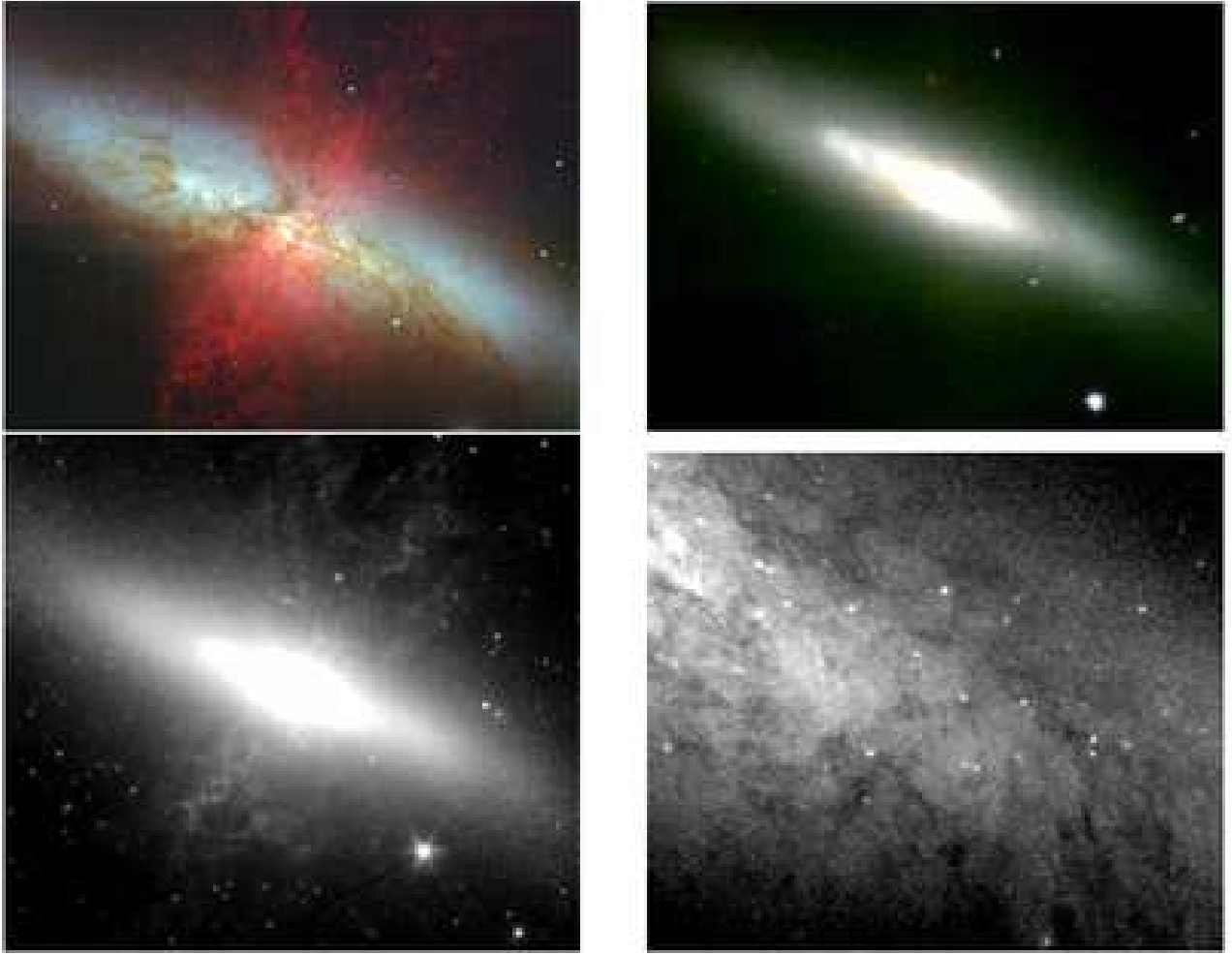}%
  \caption{Multi-band views of M82 illustrating the interplay between
stars, gas and dust. In the color composite optical (BV and H$\alpha$) 
Subaru image (top-left), we see the stellar disk and the dust lanes,
with the H$\alpha$ emission tracing a bi-conical structure along the
minor axis. In the NIR ($JHK$; top-right), and Spitzer 3.6$\mu$m 
(bottom-left) images, the most prominent structure is the central bar, 
which is completely obscured in the optical image. Filaments along
the minor axis of the galaxy in the Spitzer image, trace the distribution
of PAH particles and hot dust. 
A zoom of the SW part of the HST I-band image is shown in the bottom-right 
panel. Notice the grainy appearance of the image, which is due to the fact 
that individual stars and compact star clusters are resolved in this image.
}
  \label{fig:m82image_opt_nir}
\end{figure*}
\section{The central kiloparsec} 

\subsection{The nuclear starburst and its large-scale influence}

The nuclear starburst occupies a region of around 500~pc in size, and harbors
around $6\times10^8$~\msun\ of mass in stars \citep{For01}. 
Starburst activity in the nucleus
is going on in the form of intermittent bursts for the last 30~Myr, 
with two prominent bursts occurring in the last 10~Myr. Most of the stars 
formed in the bursts are concentrated in more than 200 compact clusters,
popularly known as Super Star Clusters\footnote{It is interesting to note 
that \citet{Carr86} had identified these compact objects on the 
photographic images and called them Superclusters almost a decade before 
they were discovered by the Hubble Space Telescope (HST) and eventually 
called by the same name.}
\citep[SSCs]{Oco95, May08}.
The energy liberated from massive stars in these clusters is released 
as galactic superwinds, which are traced by the H$\alpha$ and X-ray emitting gas
as cone-shaped structures along the minor axis of the galaxy up to distances
as far as 10~kpc above the galactic plane \citep{Lehn99}. 
The central starburst region has a patchy appearance in the optical bands
with more than 4 mag of visual extinction if the majority of the obscuring 
dust resides in a foreground screen and 43--52 mag if dust is mixed with the 
stars\citep{For01}.
Surrounding the active starburst region, there is a molecular ring of 400~pc 
radius \citep{Shen95}. The patchy optical appearance and the 
H$\alpha$ emitting cone are illustrated as a (color) composite image,
formed using the $B,V$ and H$\alpha$ images taken with Subaru telescope, 
in the top-left panel of Figure~\ref{fig:m82image_opt_nir} \citep{Ohy02}. 

The nature of the off-planar structures seen in the H$\alpha$ images were 
historically argued to be of explosive origin \citep{San64}. However, 
polarization observations by \citet{Vis72}, \citet{Sch76} and \citet{Sca91} 
established that much of the 
halo H$\alpha$ emission is in fact the disk light scattered by the dust 
particles in the halo. \citet{Bla88} and more recently \citet{Ohy02} 
established that it constitutes
two components: the first component consists of filamentary, 
narrow-line emission (full width at half maximum (FWHM) $<200$~km\,s$^{-1}$) 
organized into a bipolar outflow over the surfaces of two elongated bubbles,
and the second component consists of broad-line (at FWHM 300~km\,s$^{-1}$) and 
continuum emission arising from a faint exponential halo that rotates 
slowly with the disk.
The mid infrared spectra of the halo taken by 
recent observations by the Spitzer telescope have confirmed the presence
of dust in the halo as is illustrated in the bottom-left panel of 
Figure~\ref{fig:m82image_opt_nir} \citep{Eng06}. 

\subsection{The central galactic components: The Bar and the Bulge}

The almost dust-free vision of M82 in the near infrared bands, allowed 
\citet{Tele91} to discover a bar of $\sim$1~kpc length as soon as the imaging
technology became available at these wavelengths.
The kinematical data are fully consistent with the stars orbiting in radial
orbits along this bar \citep{Will00}. The bar contributes substantially to the 
dynamical mass in the central region. \citet{Sofu98} estimated a 
mass of $\sim10^{10}$\,\msun\ for the entire galaxy, with most of this mass
concentrated within the central 2~kpc.
A near infrared (NIR) composite image from \citet{May05} is 
presented in the 
right panel of Figure~\ref{fig:m82image_opt_nir}, where it can be clearly 
seen that the bar dominates over a smooth disk component at these wavelengths.

How big is M82's bulge? \citet{Sofu98} interpreted the entire galaxy
as the bulge of what was once a normal late type galaxy. 
The near infrared intensity profiles (see Figure~\ref{fig:m82prof})
show an excess nuclear light above that expected for an exponential disk,
which was probably interpreted as coming from the bulge component 
by \citet{Sofu98}. However, a detailed
surface photometric analysis of the profiles suggest that the observed
excess nuclear light can be completely explained in terms of the contribution 
from the nucleus and the bar. 
\citet{Gaff93} have looked for kinematical evidence for the bulge and gave an
upper limit of $10^7$~\msun\ for the bulge mass with a maximum size of 7.5~pc.

\section{The inner (1--3~kpc) disk: Morphology and dynamics}

\subsection{Optical and near infrared surface photometry}

\begin{figure}[!t]
  \includegraphics[width=\columnwidth]{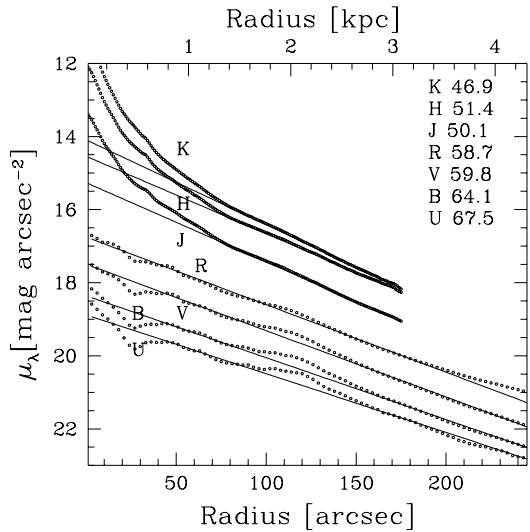}
\vspace*{-12mm}
  \caption{Multi-band radial surface brightness profiles of M82.
The numbers at the top-right corner correspond to the exponential
scale length in arcseconds in the indicated bands.
Notice the gradual flattening of the profiles at shorter wavelengths.
The bar is responsible for the steep increase of intensities 
in the NIR bands in the central part. 
}
  \label{fig:m82prof}
\end{figure}

Detailed optical surface photometric analysis of M82 is hindered by 
the obscuration caused by the dust along the line of sight, which is 
expected given the large inclination of the disk of the galaxy to the
sky-plane (77$^\circ$). 
Disk outside the central starburst suffers considerably lower 
extinction --- nevertheless the obscuring dust can be easily traced in 
the optical images even at 2~kpc ($\sim120\arcsec$) distance away from the
nucleus. On the other hand, the NIR surface brightness is not affected by
the dust. These are illustrated in Figure~\ref{fig:m82image_opt_nir}. 

\citet{Bla79} made a first attempt to obtain a detailed surface photometry
of the disk of M82, using the photographic and electronographic images
in the $B$ and $V$-bands. They found that the profiles for galactocentric 
distances $R_{\rm G}<90\arcsec$  and $R_{\rm G}>210\arcsec$ are 
indistinguishable from those of normal, late type spiral galaxies. Between 
these radii there is evidence for an enhancement which can be attributed to 
scattering of light from the inner component or disk by dust particles.
We confirm their findings using new digital data (Figure~\ref{fig:m82prof}), 
where we
can clearly see the enhancement as a bump in the radial intensity profiles
between $R_{\rm G}\sim$90--150$\arcsec$. The bump is stronger at shorter
wavelengths, suggesting that the excess light is relatively bluer --- a
characteristic property of scattered light. It can also be noticed in 
this figure,
that the profiles at $R_{\rm G}>60\arcsec$ can be very well fitted by an
exponential function, with the scale-lengths systematically increasing
at shorter wavelengths --- again a property characteristic of dusty disks 
\citep{Evan94}.
The shortest scale-length is reached in the $K$-band, where the observed
scale-length (47$\arcsec=0.8$~kpc) is expected to represent the 
scale-length of the mass distribution.
In the very central part, there is clearly an excess of NIR light
over the exponential disk, which is associated with the central
starburst and the bar, first noticed by \citet{Tele91}.

It may be recalled that the ``failure to resolve individual stars or 
stellar complexes'' was one of the defining criteria for the Irr II 
(or amorphous) morphological class assigned to M82.
The amorphous appearance seen on the ground based images of M82, breaks
down at the resolution of HST/ACS images \citep{Mut07}, especially in 
the F814 (I) band
as can be seen in the bottom-right panel of Figure~\ref{fig:m82image_opt_nir}. 
At mid infrared wavelengths, Spitzer images show the graininess even
at arcsec resolutions (bottom-left panel of 
Figure~\ref{fig:m82image_opt_nir}). The amorphous appearance on the 
ground-based optical images is caused by the dust scattering which acts 
to smoothen the images. The contribution of the scattered light progressively 
decreases at longer wavelengths, paving the way to visualize the 
inherent morphological structures and individual stars.

\subsection{The spiral arms and the Irr II classification}

The classification of Irr II does not contradict the presence of spiral
arms, one of the clearest examples being NGC972 \citep{May98}.
\citet{Oco78} have noted that the gross properties of
M\,82 --- its mass, luminosity, size, mean spectral type --- are comparable
to those of normal late type (Sc/Irr) galaxies.
\citet{Bla79}, based on the radial intensity profile 
analysis, suggested that the overall structure of M82 
resembled that of late-type spiral galaxies, rather than an irregular galaxy. 
The presence of spiral arms were always suspected 
in this amorphous galaxy: N-body simulations of the interaction
of this galaxy with the members of M81 group predicted the generation
of spiral arms \citep{Yun99}; 
feeding of the nuclear starburst suggested the presence 
of spiral arms, without which it is difficult to explain the transport of gas 
from several kiloparsec distances to all the way up to the end of the bar
\citep{Barn92}. 

\begin{figure}[!t]
  \includegraphics[width=\columnwidth]{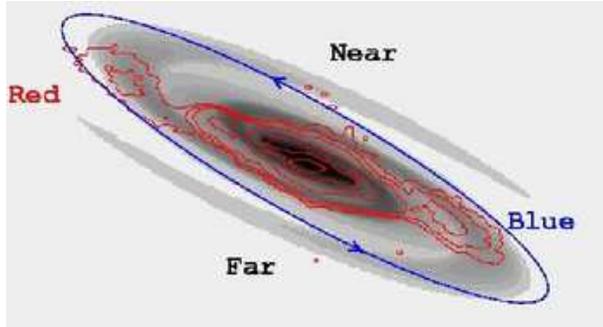}
  \caption{Illustration of the spiral arms of M82. The gray-scale 
image shows the spiral arms that best match the $K$-band residual
image (contours), which was obtained by subtracting an exponential disk from 
the observed image. The ellipse with arrows indicates the sense of 
rotation of stars expected for trailing spiral arms.
}
  \label{fig:m82spiral}
\end{figure}

\citet{May05} discovered the spiral arms in M82 by analyzing deep $JHK$-band 
images, and using the unsharp masking technique --- a method popularized by 
David Malin for amplification of faint photographic images \citep{Mali78}. 
In Figure~\ref{fig:m82spiral}, we display the modeled image of the spiral arms.
The absence of a dominating bulge, and 
the relatively open arms (pitch angle=14$^\circ$) suggest that M82 
is a late-type spiral (SBc), which is consistent with the gross properties
discussed above. 

In normal galaxies, the spiral arms are easily traced in the H$\alpha$ and
the $B$-band images due to the preferred association of the star forming
regions with 
the arms. We find that the arms of M82 are not forming stars at present.
In fact, the entire disk of M82 has stopped forming stars at around
100~Myr ago, as will be discussed in \S4. In the absence of star-forming
HII regions, the discovery of arms requires one to trace the underlying
stellar continuum emission in the optical and NIR wavelengths.
However, (1) the high obscuration caused by dust, 
(2) decreased contrast of the arms due to the scattered light, and 
(3) the nearly edge-on orientation of the disk 
make the detection of the spiral arms in the optical images an 
impossible task. The first two effects can be partly overcome
by using the NIR images. However, NIR band has a dis-advantage for
searching spiral arms --- spirals are generally bluer than the surroundings
making the spirals intrinsically less prominent at the NIR wavelengths.
Thus, it is not surprising that the spiral arms could be detected only after
subtracting the smooth exponential component even in the NIR images. 

\subsection{The gas and dust content}

Late-type spiral galaxies are rich in gas content, with the neutral gas
extending all the way up to, and some times even beyond, the optical disk.
M82 as a galaxy is abnormally gas-rich with the gas fraction by mass being 
as high as 30--40\% \citep{Youn84}.
However, most of the disk gas is concentrated in the central 1~kpc region.
There exists gas outside this radius but most of it is 
warped southwards in the direction of its companion M81 \citep{Yun93,Walt02}. 
Thus much of the disk of M82 is gas-poor.

Almost all of the H$\alpha$ emission lies in the nucleus and the halo,
with HII regions not even detectable in the spiral arms. 
However, there is ample evidence
for the presence of dust in the inner disk. These evidences are:
(1) spectra all the way to 3~kpc show Na absorption lines, whose
strength and line width suggest their interstellar origin \citep{Gotz90},
(2) spectral energy distribution of the disk obtained from long-slit
spectra and multi-band photometry suggests a visual 
extinction of $\sim$1~mag (see the contribution of Rodriguez-Merino et al. 
in this volume), and 
(3) the cluster colors suggest the presence of extinguishing dust
at parsec scales \citep{May08}. 

\subsection{The mass and mass-to-light ratio}

The rotation curve of M\,82 has been derived in several studies using both
stellar and gaseous tracers up to a radius of 170\arcsec. 
\citet{Maya60}
and \citet{Gotz90} obtained a rotation curve using optical emission and
absorption lines, whereas \citet{Sofu98} used the CO and H\,I lines.
In all these studies, the rotation curve has been modeled to obtain the radial
distribution of mass. In the radial zone between 70--170\arcsec, 
\citet{Maya60} estimated a mass of $4\times10^9$\msun,
whereas the masses estimated by \citet{Gotz90} are $\sim50$\% lower.
\citet{Sofu98} found that most of the mass is concentrated within the
central 1~kpc radius, with the mass distribution outside this radius
consistent with an exponential mass surface density profile.
Significantly, there is no evidence for a dark matter halo, and hence
the entire mass outside the central bar can be associated with the stars
and gas in the disk. 

\begin{figure}[!t]
  \includegraphics[width=\columnwidth]{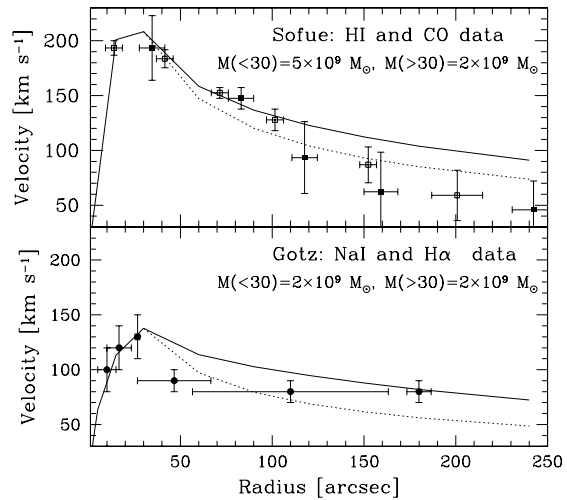}
\vspace*{-13mm}
  \caption{The rotation curve and mass model for M82 using cold gas (top) 
and stellar and ISM absorption line (bottom) velocity measurements. 
The vertical error
bars denote the errors in the measured velocities. The horizontal
error bars denote the beam size in the top panel and the range of radius
where the measured velocities remain almost the same, in the bottom panel.
The dotted line corresponds to the
rotation curve for a mass distribution that is entirely concentrated within
the 30\arcsec\ radius (solid body inside, and Keplerian outside this radius).
The effect of adding an exponential disk of scale length 48\arcsec\ is shown
by the solid line.
}
  \label{fig:m82mass}
\end{figure}

We compiled all the reliable velocity data from the
literature and re-calculated the maximum mass that the data permit outside the
nuclear starburst zone for an exponential mass distribution of scale-length
48$\arcsec$, the value derived from the $K$-band image. The resulting fits
to the observed rotation curve are shown in Figure~\ref{fig:m82mass}. 
It can be seen that the peak of the rotation curve is reached at a radius
of 30\arcsec, with the disk mass outside this radius contributing very 
little to the observed velocities. The peak velocity measured using the
cold gas tracers (HI and CO by \citet{Sofu98}) is substantially higher than 
the optical tracers, with the result that the inferred mass in the starburst 
region is a factor of 2.5 higher for the former.
However, both the tracers give a maximum mass of $2\times10^9$~\msun\ 
outside this zone, which is consistent with the values derived 
by \citet{Gotz90}. 
Given that both the HI and CO have been detected in the cone of gas
that is outflowing from the nuclear zone, and also along the off-planar
streamers, the velocities measured by these tracers may not represent
the circular velocities in the plane. Hence, velocities measured by the
optical tracers, especially the stellar absorption lines such as the
Balmer lines, are expected to be more representative of the 
circular velocities.

We obtained the \mbylk\ 
of the disk outside a radius of 1~kpc by combining the disk mass
and the $K$-band luminosity in the same radial zone (70--170\arcsec).
The resulting value is $0.15\pm0.05\,{{M_{\sun}}/{L_{K\sun}}}$. 
At least 15\% of the disk mass in the radial zone outside 70\arcsec\ is
in the form of atomic and molecular gas \citep{Youn84}.
For comparison, ${{M}/{L_{K}}}=1$ if the typical stars are solar-like.
The observed low value of ${{M}/{L_{K}}}$ indicates that the disk is 
dominated by stars that are more massive than the Sun, consistent with
the A--F spectral classification assigned for this galaxy by \citet{Mor58}.

\section{The inner (1--3~kpc) disk: stellar content}

\subsection{Age of the dominant stellar population}

\begin{figure*}[!t]
  \includegraphics[width=16cm]{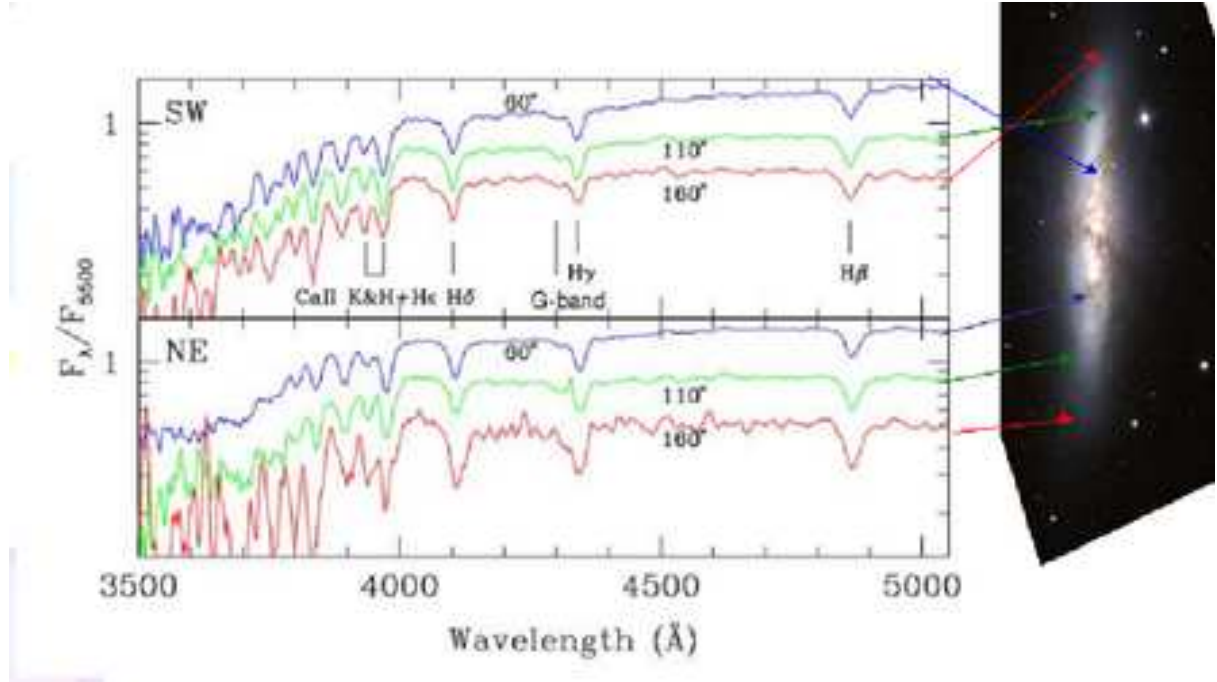}
  \caption{Blue part of the major axis spectra extracted 
on either side of the center at three radial
zones centered at galactocentric distances of 60\arcsec, 110\arcsec, and 
160\arcsec\ are shown. Approximate positions of the extracted
spectra on an optical image are shown by the arrows.
Prominent age-sensitive features are indicated.
}
  \label{fig:m82spectra}
\end{figure*}

\citet{May06} used the long-slit spectra along the major axis of M82
to age date the dominant stellar population on either side of the nucleus.
The position of slit along the major axis as well as the identification of
some of the age-sensitive absorption features are shown in 
Figure~\ref{fig:m82spectra}. 
Each spectra is obtained by averaging the long-slit spectra spatially over
a width of 40--50\arcsec. Spectrum corresponding to 60\arcsec\ radius
on the northeast side encloses the bright optical patch known as ``M82-B''. 
The four spectra corresponding to radius$\leq110$\arcsec\ are strikingly 
similar. The outer spectra show a slight
difference in the relative intensities of the Ca II K \& H lines, with 
respect to the inner disk spectra.

Comparison of the features in the observed spectra to 
those from a Single Stellar Population (SSP) model, gives an
age of 0.5~Gyr for the populations for the disk interior to 110\arcsec\ radius
(i.e. $<2$~kpc). More importantly, the spectra are not consistent with 
mean ages greater than 0.7~Gyr.
The 160\arcsec\ spectra (2--2.7~kpc radial zone) suggest a population that 
is marginally more evolved as compared to that of the inner part of the 
disk, with an age of $0.9\pm0.1$~Gyr. 

\subsection{Global Star formation history}

\begin{figure}[!t]
  \includegraphics[width=\columnwidth]{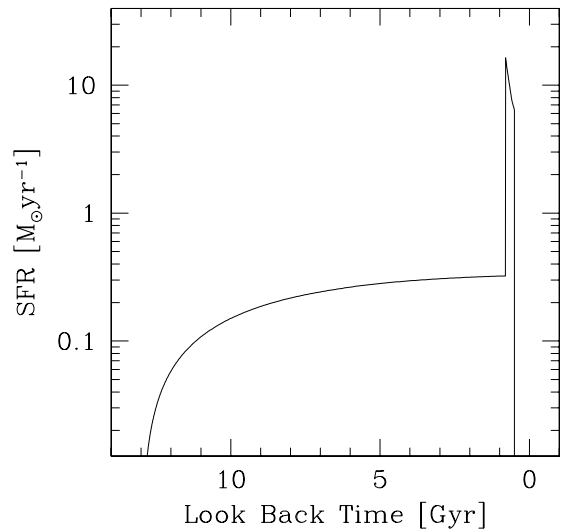}
  \caption{Star formation history of the disk of M\,82 that reproduces
best the observed properties. More than 90\% of the observed disk mass
is produced in the burst.}
  \label{fig:m82sfh}
\end{figure}

From the discussions above, it is clear that the inner disk of M82 
has a low \mbylk, and a young mean age, as compared to those for a normal 
disk of a late-type galaxy. Detailed analysis of abundances of elements
in cold, warm and neutral gas, suggests that the $\alpha$ elements
are systematically enriched \citep{Orig04}. These are unmistakable signatures 
of a disk seen immediately after a major star-burst. 
Detailed modeling of these data suggest that
almost all the observed stellar mass of the disk was part of a burst 
lasting for only a few hundred million years, the burst having completely 
stopped around 0.5~Gyr. The resulting star formation history is shown
in Figure~\ref{fig:m82sfh}.
Further details of the model can be found in \citet{May06}.

\subsection{Spatially resolved star formation history from star counting}

A recent study by \citet{Dav08a} using the IR color-magnitude
diagrams for the resolved red stars threw more light not only on the 
evolutionary status of the inner and outer disk, but also that of the halo.
In this section, we discuss the results pertinent to the inner disk, 
postponing the discussions of the results for the outer disk and halo
to the next section. Deep wide-field images taken with the 
{\it MegaCam} and {\it WIRCam} instruments on the 
Canada-France-Hawaii Telescope (CFHT)
were used in his work to trace the brightest stars seen longward of the 
traditional $R$-band. Specifically, $r^\prime$, $i^\prime$, $z^\prime$, 
$J$, $H$ and $K$-bands were used. The exposures were long enough to allow
the detection of the asymptotic giant branch stars (AGBs) in M82. 

One of the aims of Davidge's study was to test the burst nature of star
formation in the inner disk predicted by us \citep{May06}.
According to the burst model, the disk should mostly contain stars that 
are older than 0.5~Gyr. According to the isochrones of \citet{Gir02},
stars are in their AGB phase at these ages, which are above the
detection limits of the CFHT observations. Additionally,
the disk is not expected to contain stars that are formed relatively recently,
such as OB stars, and Red Super Giants (RSGs). From the analysis of 
color-magnitude diagrams (CMDs), Davidge firmly established the absence of 
these relatively
younger stars in the entire disk. He found that the brightest stars populating
the CMDs are the AGB stars. The progenitors of these stars are of a few solar
masses, with the brighter AGBs being slightly more massive and younger as
compared to the relatively fainter AGBs. 

The $i^\prime$-band luminosity function for the inner disk (2--4~kpc) 
is reproduced in the top panel of Figure~\ref{fig:m82davidge}. 
It has a peak at 
$i^\prime=23.4$~mag, the number density of stars dropping significantly
on either side of the peak. The brightness corresponding to the peak
is more than a magnitude brighter than the limiting magnitude, suggesting
that the peak is real and not due to the incompleteness on the fainter
side. Given the one-to-one relation between the age and magnitude during 
the AGB-phase, the luminosity function can be used to infer the 
star formation histories (SFHs). 
The observed luminosity function resembles exactly that expected for
a burst mode of SFH. The peak corresponds to a peak in the starburst
activity at $\sim0.5$~Gyr ago. There are hardly
any stars brighter than $i^\prime \sim22$~mag, implying the absence of
stars younger than $\sim100$~Myr.

Another finding in our study is that only less than 10\% of the presently
observed disk mass is contributed by the stars older than about 1~Gyr.
Otherwise, the disk stellar mass would exceed the observed dynamical mass 
for the inner disk. Data give an indication of a decrease in the number 
density of stars older than $\sim1$~Gyr. However, the detection limit and the 
present status of understanding of the evolution of AGB stars, do not allow 
a conclusive answer about the absence of these older stars.

Thus, star formation history derived from the resolved stellar population
is in excellent agreement with the burst model we had proposed. These
new observations can be used to fine tune the burst model. It seems the burst
strength decreased gradually after it peaked at $\sim0.5$~Gyr, continuing
up to 100~Myr ago, terminating abruptly at this age.
The continuation of the burst to relatively younger ages as compared to
our model, implies that the \mbylk\ could be accommodated with slightly
more mass in the pre-burst stars. Thus the mass in these stars could be
slightly higher than the 10\% of the disk mass predicted by us.

\section{The Outer Disk and The Halo}

\subsection{The outer stellar disk}

The short scale length of the M82 disk implies that the disk intensity falls
more rapidly with radius in this galaxy as compared to that in normal 
late-type galaxies. As a consequence most of the detailed information on the 
disk exists only for the inner 4~kpc. The disk size defined as the 
radius at which $B$-band surface brightness falls to 25~mag\,arcsec$^{-2}$
is as big as 6.5$^\prime$ (6.9~kpc). Thus the discussions of the previous
sections pertained to basically the inner half of the disk. 
From now-on-wards, we refer to the part of the disk at galactocentric
distances greater than 4~kpc as the outer disk.

Did the outer disk also participate in the large-scale starburst
following the interaction? Does it also lack stars older than the burst?
Obtaining spectra that could be used to age date the outer 
disk ($\mu_B>22$~mag\,arcsec$^{-2}$) is challenging even for the largest 
present-day telescopes. In the absence of spectroscopic ages, approximate
ages of the prominent stellar population could be estimated using the
mass-to-light ratios. However, the outer disk lacks a commonly used 
tracer of rotation curve such as H$\alpha$, or HI, and hence the dynamical 
mass estimates are not available.

\begin{figure}[!t]
  \includegraphics[width=\columnwidth]{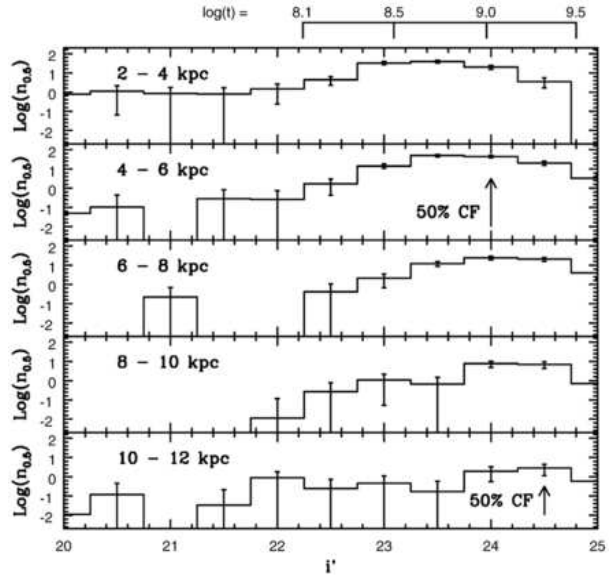}
  \caption{
The $i^\prime$-band Luminosity Functions (LFs) of stars in the M82 disk. 
$n_{0.5}$ is the number of stars with $r^\prime - i^\prime$ between 0 and 2 
per 0.5 $i^\prime$ mag per arcmin$^2$. See \citet{Dav08a} for more details.
}
  \label{fig:m82davidge}
\end{figure}

The CMD of resolved stars in the NIR bands offers an alternative way to
determine the ages of stellar populations and SFH of the outer disk.
Using such a technique, \citet{Dav08a} found that the AGB stars are the
brightest NIR stars in the outer disk, similar to the results found
for the inner disk. AGB stars are detected even 
at radial distances as far out as 12~kpc from the center. 
Throughout the outer disk, there is a clear lack of stars younger 
than $\sim100$~Myr, which suggest complete cessation of star formation
activity in the last $100$~Myr. 
The $i^\prime$-band LF for three radial zones are displayed in the 
bottom three panels of Figure~\ref{fig:m82davidge}.
It can be seen the SFH of the outer disk is qualitatively very similar
to that of the inner disk. In the figure, we can see a trend of the peak 
of the LF shifting systematically to fainter magnitudes at larger radius.
The existence of a peak before the 50\% confusion limit
suggests that the star formation in these parts of the disk also occurred
in a burst mode, and that the observed trend suggests 
that the most active period of star formation was systematically earlier
farther out in the disk. At around 10~kpc distance from the center, the burst
seems to have started around 3~Gyr ago.

The above ages should be considered tentative, until they are corroborated 
with observations that go around 1 or 2 magnitudes deeper. If they are found 
to be correct, then it calls for a new modeling regarding the triggering 
of the burst. M82 went through an interaction with the members of M81 group 
around 500~Myr ago. The age of the burst in the inner disk is consistent 
with the idea that the interaction was responsible for its triggering.
However, if the bulk of the red population is as old as 3~Gyr,
then clearly the burst couldn't have been triggered by the latest interaction.

\subsection{The stellar halo}

\citet{Saka99} used the HST/WFPC2 $V$ and $I$-band images to study the 
properties of the resolved stellar populations of two adjacent fields in the
northeast part of the halo. The limiting magnitude in these fields
was around 1 magnitude below that expected for the tip of the red giant branch
(TRGB) stars, and the
crowding of stars was not a serious problem in defining the TRGB. 
In both the fields, they detected large quantities of RGB stars, and 
also candidates for AGB stars. One of their fields passed through the
off-planar HI filament, where they found significantly higher fraction of
AGB stars, indicating the presence of intermediate age populations.
\citet{Dav04} studied a region in the halo at a projected distance of 
$\sim$1~kpc south of the M82 disk plane, using the $H$ and $K'$-band
images at 0.08$\arcsec$ FWHM and found conclusive evidence for the
presence AGB stars.

In a more recent work, 
\citet{Dav08a} studied the population residing in the halo along the
minor axis of the galaxy. 
AGB stars were detected even at 7~kpc distances about the disk plane.
But unlike in the disk, stars of younger ages are also found in the
halo. Davidge argued that the majority of halo stars are formed in situ
in small associations or clusters and then dispersed to form the
diffusely distributed population in the halo.
It may be noted that there are several isolated Orion-like star-forming 
regions in the intra-group medium of M81 group \citep{deM08},
including one $\sim6$~kpc to the south of M82 \citep{Dav08b}.

\citet{Sait05} had reported the discovery 
of two globular clusters at the extreme south of the galaxy. \citet{Dav08a}
also encountered 5 objects that can be candidates for globular
clusters. However, all these objects lie in the direction of M81, putting
some doubts regarding their affiliation to M82.

\section{The Super star cluster population in the post-starburst disk}

One of the biggest achievements of the HST is the discovery of the blue 
compact star clusters in external galaxies, frequently referred as Super 
Star Clusters or SSCs. Starburst activity seems to be the 
necessity condition for their formation and the starburst nucleus of M82 is 
one of the first regions where SSCs in large numbers are found \citep{Oco95}.
If the disk of M82 suffered a large-scale starburst following its interaction
with the members of M81 group, then it is very likely
that several SSCs were formed, and are still surviving.
The wide field imaging capability of the ACS provided an opportunity to
search for the SSCs in the disk of M82. 
A rich population of clusters is indeed discovered, some of which can be
easily noticed in the bottom-right Figure~\ref{fig:m82image_opt_nir} 
(bright points).
The details of the search and their properties including their evolutionary
status are published in \citet{May08}. 
Here, we discuss the properties of the cluster population from the point
of view of M82 as a galaxy.

\begin{figure}[!t]
  \includegraphics[width=\columnwidth]{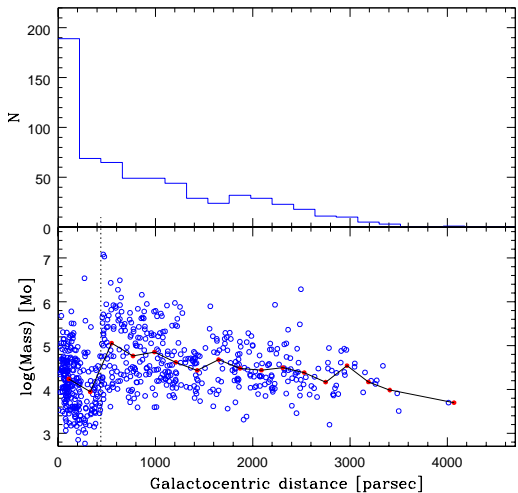}
  \caption{Radial distribution of the number (top) and mass (bottom) of the 
Super Star cluster population in M82. The mean mass of the cluster
population at each radius is shown by the solid line.
The dotted vertical line in the bottom panel divides the nuclear clusters
from the disk ones.
}
  \label{fig:m82av}
\end{figure}

A total of 653 clusters were found, nearly 400 of these situated outside 
the nuclear zone. Spectroscopic ages are available only for a handful
of these clusters, with none of the clusters older than 200~Myr \citep{Kon08}.
Given that the disk is not heavily affected by the obscuration by dust,
optical colors can be used for estimating the ages of the clusters. Both
the $V-I$ and $B-V$ colors are consistent with an age of less than 0.5~Gyr, 
with the visual extinction of around 1~mag. 
The radial distribution of the number and derived mass of the clusters
is given in Figure~\ref{fig:m82av}. The cluster density is maximum within
200~pc of the nucleus, thereafter decreasing almost monotonically.
The mean as well as the maximum mass of clusters are the highest 
at around 0.5~kpc distance from the nucleus, just where the bar ends. 
The mean cluster mass decreases as a function of its distance from the center.

There are no detectable clusters at galactocentric distances greater than 
4~kpc. The ACS images cover around 6~kpc distance along the major axis 
on either side of the nucleus. In addition, given the low surface brightness 
of the outer disk, the detection limit is more than a magnitude fainter than
in the inner disk. Hence, the non-detection of compact clusters in the outer 
disk is intriguing. 

The absence of SSCs in the outer disk could be either because the
conditions were not suitable for the formation of SSCs
or that the SSCs there are preferentially destroyed. On an average
SSCs are less massive and more extended at larger galactocentric distances, 
making them more and more vulnerable to the destruction by tidal forces.
Thus, even if SSCs were formed in the outer disk, they could have been
destroyed.

\section{M82 as a galaxy}

The low inferred mass, absence of a dominating bulge and relatively open 
nature of the newly discovered spiral arms all point to a late morphological 
type for M82. The relative scarcity of globular clusters is also consistent
with a late morphological type.
\citet{Dav08a} compared the stellar properties of the disk of M82 with 
NGC2403, a late type isolated galaxy with mass similar to that of M82.
He found that M82's stellar disk extends at least up to 12~kpc, which 
is comparable to the size of the stellar disk of NGC2403 ($\sim14$~kpc), 
inferred using the same technique as in M82.
He also found that the number of bright AGB stars per unit surface brightness
(in the $K$-band) in the outer disks of M82 and NGC2403 are similar.

However, M\,82 differs from normal spiral galaxies in the following ways:
disks of normal spiral galaxies are characterized by star formation
that is continuing uninterrupted over the Hubble time, with mean 
stellar ages of $>5$~Gyr \citep{1994ApJ...435...22K}. 
On the contrary, there are no stars formed in the recent 100~Myr in the
disk of M82 and most of the disk stars are formed in a burst with
mean age as young as 0.5~Gyr. This peculiar SFH makes M82 deviate 
from the normal galaxies in the relations between
global parameters: it is too bright and also metal rich for its relatively
low estimated mass of $\sim10^{10}$~\msun\ \citep{2004AJ....127.2031K}.

It is interesting to note that a number of interacting galaxies also
show disk-wide star formation at present: examples being NGC\,4038/9,
Arp\,299 and NGC\,4676 \citep{1998A&A...333L...1M, 1999AJ....118..162H,
2000ApJ...541..644X}. This kind of widespread star formation in galactic disks
is probably induced by shocks generated in the interstellar medium following
the interaction \citep{2004MNRAS.350..798B}.

If the disk-wide starburst happened in the existing stellar disk, then
we should be able to infer the presence of such a disk through observations
of the pre-burst stars. As of now there is no compelling support for the 
existence of such a disk. This gives rise to two possibilities:
(1) the pre-collisional disk was intrinsically of low surface brightness (LSB),
 or (2) it was a normal late-type galaxy, but its disk, along
with the metal-enriched gas was stripped during the collision.
Alternatively, M82 could be in the process of formation, with the starburst
being the first major event of star formation in a purely gas-rich galaxy.
We discuss these three cases individually below.

\subsection{Gas accretion and starburst in an LSB galaxy}

\citet{Miho99} discussed pre-collisional galaxy properties that favors
star formation in the disk, before the burst is turned on in the center.
He found that the pre-collisional galaxy should be a normal high-surface
brightness galaxy with a big bulge or a gas-rich LSB galaxy.
If the pre-collisional M\,82 had a big bulge, it would have survived the
interaction \citep{Sofu98}. However, the observed bulge of M\,82 is small,
consequently, pre-collisional M\,82 should have been a gas-rich LSB galaxy.
Even if the LSB galaxy was gas-poor to begin with, it might have acquired 
vast amounts of gas as it drifted through the clouds in the intra-group 
medium of M81 as suggested by \citet{Youn84}. 
Deep observations aimed at detecting a stellar disk with properties
expected for known LSBs would be required to test this scenario.

\subsection{Rejuvenated star formation in a stripped stellar disk}

In the scenario proposed by \citet{Sofu98},
a disk of old stars is not expected underneath the newly formed disk.
Instead, the old stars from the stripped stellar disk and the halo
should be found somewhere in the M81 group. In recent years
several fields in the intergalactic medium in the M81 group had been 
the target of HST observations. However, none of these studies so far 
have reported the discovery of stripped stars. It should be noted that
the majority of the pre-collisional stars are expected to be low-mass 
main sequence stars, which are below the detection limit of the HST
observations. The brightest pre-collisional stars are expected to be
RGB stars, which are also not detected in the intra-group medium. 
\citet{Saka99} detected the RGB stars in the halo at a radial
distance of 2~kpc. The stripped stars are expected to be much farther
than this, else they would have contributed to the rotation curve.

However, not all the available data today favor this hypothesis. 
For example, according to this hypothesis stellar content and 
velocity field of M82 should be 
dominated by those typical of bulge stars. On the other hand, the stars
dominating the light immediately outside the nuclear region are much
younger than that in typical bulges, and the velocity fields are dominated
by the bar potential rather than the bulge \citep{Will00}.

\subsection{A nearby galaxy in formation?}

M\,82 resembles the luminous blue compact galaxies seen at 
redshifts 0.2--1.0 in having high luminosity, short scale length,
a nuclear starburst and low mass \citep{2006ApJ...640L.143N}. 
Thus in M\,82, we may be
witnessing stars being formed for the first time in its disk, probably
in the debris left behind in the interaction. In other words, M\,82
could be a nearby galaxy in formation. If this is the case, then the 
detection of RGB stars in the halo of M82 may have been acquired 
recently.

While we can not rule out completely none of the three models proposed here,
majority of the observations favor the starburst in a gas-rich LSB model.

\section{Summary and Future Research}

Studies in the last decade have undoubtedly established that M82 is a late 
type spiral galaxy. Its unusual optical appearance, which was responsible
for its distinct historical classification, 
is the combined effect of dust scattering, high inclination, and peculiar
star-formation history. The disk of the galaxy formed most of its stars 
in a burst mode, 
rather than continuously over the Hubble time like in normal late type 
galaxies. The disk-wide burst model was initially proposed based on the data 
on the inner disk (radius$<3$~kpc). However, recent data on the outer parts 
suggest that the burst model is applicable as far out in the disk as 
12~kpc radius. The star-formation in the disk was completely quenched
following the burst, as evidenced from the lack of 
stars younger than $\sim100$~Myr anywhere in the disk.
The data seem to suggest that the starburst activity
stopped systematically at earlier times at distances farther out in the
disk. It may be possible that the starburst activity first started in the
outer disk, which then moved progressively to inner radii. 

It is now well established that the burst activity of the inner disk is related
to the latest encounter between M82 and its neighbors M81 and NGC3077.
According to the scenario proposed by \citet{Sofu98}, the
interaction happened around 1~Gyr ago. Numerical simulations of \citet{Yun99}
indicate that the last encounter happened more recently, around 300~Myr ago. 
In the burst model that best fits the observational data for the inner disk, 
the burst started around 800~Myr ago, and lasted for around 300~Myr.
The duration, as well as the commencement of the burst were mainly constrained
by the observed \mbylk, which may be 
uncertain by a factor of two. If we take into account this uncertainty, data 
can be reconciled with younger burst ages.
Thus the stellar population ages in the inner disk are in general 
consistent with the idea that the burst was triggered by the interaction.
The same is not true for the outer disk, where the 
bulk of the population seems to be as old as 1~Gyr. At distances
10~kpc away from the center, majority of the AGB stars seems to be as old 
as 3~Gyr. So if the star formation started in the outer disk, and
subsequently propagated inwards, then the burst should have initiated 
before the latest encounter. 
Was the encounter happened much before the present estimates 
or was the starburst related to any earlier encounter?

The pre-burst stellar disk of M82 most likely resembled that of
a low surface brightness galaxy. Detection of such a disk
using the traditional surface photometric techniques is 
difficult due to the presence of the glare of the burst stars.
Identification of the pre-burst stellar population in the CMD would be 
the surest way for testing the existence of the pre-collisional disk.
These stars are expected to be occupying the 
main sequence and red giant branches of the CMD. The detection limit in the
study of \citet{Dav08a} was not deep enough to identify these stars.
The limiting magnitude
for point sources of HST/ACS observations of M82 is above the main
sequence turn-off magnitude, but are well below the tip of the RGBs.
Hence, these images especially at radius greater than 2~kpc could be used
for the detection of RGB stars. As of now, such a study was carried out 
only for the halo, resulting in the discovery of RGB stars \citep{Saka99}.
Detection of RGB stars in the disk will not only put strong constraints
on the mass of the stars formed before the starburst, but also would help
in understanding the evolutionary state of M82.

\section{\normalsize Acknowledgements.}
We thank Alessandro Bressan and Daniel Rosa for discussions on a variety 
of topics discussed in this review. We acknowledge the contribution of
Lino Rodriguez, Abraham Luna and Roberto Romano in some part of the work
presented here. Finally, it is a pleasure
to thank my (YDM) co-author in all the works related to M82. Thank you
very much Luis. It is you who motivated me to work on the NIR astronomy and M82.


\begin{thebibliography}

\bibitem[Barnes (2004)]{2004MNRAS.350..798B}
Barnes, J.~E.\ 2004, \mnras, 350, 798

\bibitem[Barnes \& Hernquist (1992)]{Barn92}
Barnes, J. E. \& Hernquist, L. 1992, \araa, 30, 705

\bibitem[Blackman, Axon \& Taylor (1979)]{Bla79}
Blackman, C.P., Axon, D.J. \& Taylor, K. 1979, MNRAS, 189, 751

\bibitem[Bland \& Tully(1988)]{Bla88} 
Bland, J., \& Tully, B.\ 1988, \nat, 334, 43 

\bibitem[Carrasco et al. (1986)]{Carr86}
Carrasco, L., Recillas-Cruz, E., Cruz-Gonzalez, I. \& Melnick, J. 1986,
Rev. Mexicana Astron. Astrof. 12, 135

\bibitem[Davidge (2008a)]{Dav08a}   
Davidge, T.J. 2008, AJ, 136, 2502

\bibitem[Davidge (2008b)]{Dav08b}   
Davidge, T.J. 2008, \apj, 678, L85 

\bibitem[Davidge et al.(2004)]{Dav04} 
Davidge, T.~J., Stoesz, 
J., Rigaut, F., Veran, J.-P., \& Herriot, G.\ 2004, \pasp, 116, 1 

\bibitem[de Mello et al.(2008)]{deM08} 
de Mello, D.~F., Smith, L.~J., Sabbi, E., Gallagher, J.~S., Mountain, M., 
\& Harbeck, D.~R.\ 2008, \aj, 135, 548 

\bibitem[de Vaucouleurs (1959)]{deVa59}
de Vaucouleurs, G. 1959, Handbuch der Physik, 53, 275

\bibitem[Engelbracht et al. (2006)]{Eng06}
Engelbracht et al. 2006, \apj, 642, L127

\bibitem[Evans (1994)]{Evan94}
Evans, R. 1994, MNRAS, 266, 511

\bibitem[F{\"o}rster Schreiber et al.(2001)]{For01} 
F{\"o}rster Schreiber, N.~M., Genzel, R., Lutz, D., Kunze, D., 
\& Sternberg, A.\ 2001, \apj, 552, 544 

\bibitem[Freedman et al. (1994)]{Free94}
Freedman, W. L. et al. 1994, \apj, 427, 628

\bibitem[Gaffney, Lester \& Telesco (1993)]{Gaff93}
Gaffney, N. I, Lester, D. F,  \& Telesco, C. M. 1993, \apj, 407, L57

\bibitem[Girardi et al.(2002)]{Gir02} 
Girardi, L. et al.  2002, \aap, 391, 195

\bibitem[Goetz et al.(1990)]{Gotz90}
Goetz, M., McKeith, C.D., Downes, D., \& Greve, A. 1990, \aap, 240, 52

\bibitem[Gottesman \& Weliachew (1977)]{Gott77}
Gottesman, S. T., \& Weliachew, L. 1977, \apj, 211, L57

\bibitem[Hibbard \& Yun(1999)]{1999AJ....118..162H}
Hibbard, J.~E., \& Yun, M.~S.\ 1999, \aj, 118, 162

\bibitem[Holmberg (1950)]{Holm50}
Holmberg, E. 1950, Lund Medd. Astron. Obs. Ser. II, 128, 1

\bibitem[Karachentsev et al.(2004)]{2004AJ....127.2031K}
Karachentsev, I.~D., Karachentseva, V.~E., Huchtmeier, W.~K., \&
Makarov, D.~I.\ 2004, \aj, 127, 2031

\bibitem[Kennicutt et al.(1994)]{1994ApJ...435...22K}
Kennicutt, R.~C., Tamblyn, P., \& Congdon, C.~E.\ 1994, \apj, 435, 22

\bibitem[Konstantopoulos et al.(2008)]{Kon08} 
Konstantopoulos, I.~S., Bastian, N., Smith, L.~J., Trancho, G., 
Westmoquette, M.~S., \& Gallagher, J.~S., III 2008, \apj, 674, 846 

\bibitem[Lehnert, Heckman \& Weaver (1999)]{Lehn99}
Lehnert, M, D., Heckman, T. M., \& Weaver, K. A. 1999, \apj, 523, 575

\bibitem[Malin (1978)]{Mali78}
Malin, D. F. 1978, Nature, 276, 591

\bibitem[Mayall (1960)]{Maya60}
Mayall, N. U. 1960, Ann. d'Ap., 23, 344

\bibitem[Mayya et al.(1998)]{May98} 
Mayya, Y.~D., Ravindranath, S., \& Carrasco, L.\ 1998, \aj, 116, 1671 

\bibitem[Mayya, Carrasco \& Luna (2005)]{May05}
Mayya, Y.~D., Carrasco, L., \& Luna, A.\ 2005, \apjl, 628, L33

\bibitem[Mayya et al.(2006)]{May06} 
Mayya, Y.~D., Bressan, 
A., Carrasco, L., \& Hernandez-Martinez, L.\ 2006, \apj, 649, 172 

\bibitem[Mayya et al.(2008)]{May08} 
Mayya, Y.~D., Romano, R., Rodr{\'{\i}}guez-Merino, L.~H., Luna, A., 
Carrasco, L., \& Rosa-Gonz{\'a}lez, D.\ 2008, \apj, 679, 404

\bibitem[Mihos (1999)]{Miho99}
Mihos, J. C. 1999, in IAU Symp. 186, Galaxy Interactions at Low and High
Redshift, ed. J. E. Barnes \& D. B. Sanders (Boston: Kluwer), 205

\bibitem[Mirabel et al.(1998)]{1998A&A...333L...1M}
Mirabel, I.~F., et al.\ 1998, \aap, 333, L1

\bibitem[Morgan (1958)]{Mor58}
Morgan, W.W. 1958, \pasp, 70, 364

\bibitem[Mutchler et al.(2007)]{Mut07}
  Mutchler, M., Bond, H.~E., Christian, C.~A., Frattare, L.~M., Hamilton, F.,
         Januszewski, W., Levay, Z.~G., Mountain, M., Noll, K.~S., Royle, P.,
         Gallagher, J.~S., \& Puxley, P.\ 2007, \pasp, 119, 1

\bibitem[Noeske et al.(2006)]{2006ApJ...640L.143N}
Noeske, K.~G. et al. 2006, \apjl, 640, L143

\bibitem[O'Connell \& Mangano (1978)]{Oco78}
O'Connell, R. W., \& Mangano, J. J. 1978, \apj, 221, 62

\bibitem[O'Connell et al.(1995)]{Oco95} 
O'Connell, R.~W., Gallagher, J.~S., III, Hunter, D.~A., \& Colley, W.~N.\
1995, \apjl, 446, L1

\bibitem[Ohyama et al. (2002)]{Ohy02} Ohyama et al. 2002, PASJ 54, 891

\bibitem[Origlia et al.(2004)]{Orig04}
Origlia, L., Ranalli, P., Comastri, A., \& Maiolino, R. 2004, \apj, 606, 862

\bibitem[Rieke et al. (1980)]{Riek80}
Rieke, G. H., Lebofsky, M. J., Thompson, R. I., Low, F. J., \& Tokunaga, A. T.
1980, \apj, 238, 24

\bibitem[Sakai \& Madore (1999)]{Saka99}
Sakai, S. \& Madore, B.F. 1999, \apj, 526, 599

\bibitem[Saito et al. (2005)]{Sait05}
Saito, Y. et al. 2005, \apj, 621, 750

\bibitem[Sandage \& Brucato (1979)]{Sand79}
Sandage, A., \& Brucato, R. 1979, \aj, 84, 472

\bibitem[Sandage \& Miller (1964)]{San64}
Sandage, A. R. \& Miller, W. C. 1964, Science 144, 405

\bibitem[Scarrott et al.(1991)]{Sca91} 
Scarrott, S.~M., Eaton, N., \& Axon, D.~J.\ 1991, \mnras, 252, 12P

\bibitem[Schmidt et al.(1976)]{Sch76} 
Schmidt, G.~D., Angel, J.~R.~P., \& Cromwell, R.~H.\ 1976, \apj, 206, 888 

\bibitem[Shen \& Lo (1995)]{Shen95}
Shen, J., \& Lo, K. Y. 1995, \apjl, 445, L99

\bibitem[Sofue (1998)]{Sofu98}
Sofue, Y. 1998, \pasj, 50, 227

\bibitem[Telesco et al. (1991)]{Tele91}
Telesco, C. M., Joy, M., Dietz, K., Decher, R., \& Campins, H. 1991,
\apj, 369, 135

\bibitem[Visvanathan \& Sandage(1972)]{Vis72} 
Visvanathan, N., \& Sandage, A.\ 1972, \apj, 176, 57

\bibitem[Walter, Weiss \& Scoville (2002)]{Walt02}
Walter, F., Weiss, A., \& Scoville, N. 2002, \apjl, 580, L21

\bibitem[Wills et al. (2000)]{Will00}
Wills, K. A. et al. 2000, \mnras, 316, 33

\bibitem[Xu et al.(2000)]{2000ApJ...541..644X}
Xu, C., Gao, Y., Mazzarella,
J., Lu, N., Sulentic, J.~W., \& Domingue, D.~L.\ 2000, \apj, 541, 644

\bibitem[Young \& Scoville (1984)]{Youn84}
Young, J.S., \& Scoville, N. Z. 1984, \apj, 287, 153

\bibitem[Yun, Ho, \& Lo (1993)]{Yun93}
Yun, Min S., Ho, Paul T. P., \& Lo, K. Y. 1993, \apj, 411, 17

%
\bibitem[Yun(1999)]{Yun99} 
Yun, M.~S.\ 1999, Galaxy Interactions at Low and High Redshift, 186, 81

\end{thebibliography}
\end{document}